\documentclass[twocolumn]{aastex631}
\usepackage{amsmath}
\usepackage{natbib}

\shorttitle{UV irradiation of EA}
\shortauthors{Suhasaria et al.}
\graphicspath{{./}{figures/}}

\begin{document}

\title{Ly-$\alpha$ processing of solid-state Ethanolamine: Potential Precursors to Sugar and Peptide Derivatives}

\correspondingauthor{Tushar Suhasaria}
\email{suhasaria@mpia.de}

\author[0000-0002-0786-7307]{T. Suhasaria}
\affiliation{ Max Planck Institute f\"ur Astronomie, K\"onigstuhl 17, 69117, Heidelberg, Germany}

\author{S. M. Wee}
\affiliation{ Max Planck Institute f\"ur Astronomie, K\"onigstuhl 17, 69117, Heidelberg, Germany}

\author{R. Basalg{\`e}te}
\affiliation{Laboratory Astrophysics Group of the Max Planck Institute for Astronomy at the Friedrich Schiller University \\
Jena, Institute of Solid State Physics, Helmholtzweg 3, 07743, Jena, Germany}

\author{S. Krasnokutski}
\affiliation{Laboratory Astrophysics Group of the Max Planck Institute for Astronomy at the Friedrich Schiller University \\
Jena, Institute of Solid State Physics, Helmholtzweg 3, 07743, Jena, Germany}


\author{C. J{\"a}ger}
\affiliation{Laboratory Astrophysics Group of the Max Planck Institute for Astronomy at the Friedrich Schiller University \\
Jena, Institute of Solid State Physics, Helmholtzweg 3, 07743, Jena, Germany}

\author{K. Schwarz}
\affiliation{ Max Planck Institute f\"ur Astronomie, K\"onigstuhl 17, 69117, Heidelberg, Germany}

\author{Th. Henning}
\affiliation{ Max Planck Institute f\"ur Astronomie, K\"onigstuhl 17, 69117, Heidelberg, Germany}



\begin{abstract}

Ethanolamine (EA), a key component of phospholipids, has recently been detected in the interstellar medium within molecular clouds. To understand this observation, laboratory studies of its formation and destruction are essential and should be complemented by astrochemical models. This study investigates the photostability of EA ice under Lyman (Ly)-$\alpha$ (10.2 eV) irradiation at 10 K, and explores its potential role in the formation of simple and complex organic molecules in molecular clouds. The UV destruction cross section of EA was estimated to be ($4.7\pm0.3)\times10^{-18}$ cm$^2$, providing insight into its half-life of $6.5\times10^{7}$ yr in dense interstellar clouds. Fourier transform infrared spectroscopy and quadrupole mass spectrometry were used to identify various photoproducts, with their formation pathways discussed. Ethylene glycol and serine were tentatively detected during the warming up process following irradiation, suggesting that EA could contribute to the formation of prebiotic molecules such as sugars, peptides and their derivatives. High mass signals detected in the mass spectrometer suggest the presence of several complex organic molecules, and further analysis of residues at room temperature is planned for future work. The results suggest that EA could contribute to the formation of prebiotic molecules in space, with implications for the origin of life.

\end{abstract}


\section{Introduction} \label{sec:intro}

In the cold and dense environments of interstellar molecular clouds, gas-phase molecules accumulate on dust grains, forming ice layers. Cosmic rays penetrate these clouds, and UV photons produced by hydrogen excited by cosmic rays can enhance the chemical complexity of the existing ices. This environment promotes the formation of complex organic molecules (COMs) from simpler compounds that form on grain surfaces and evolve within the clouds. In recent years, the ever so sensitive James Webb Space Telescope has helped to observe the chemical complexity in these regions in the solid state \citep{mcclure2023, chen2024, rocha2024}.

Recently, ethanolamine (NH$_2$CH$_2$CH$_2$OH, or EA) was detected in the interstellar medium within the molecular cloud G+0.693-0.027 in the Sagittarius B2 complex at the Galactic Center. Reported by \citet{rivilla2021}, this detection was achieved through rotational transition observations and suggests that EA may form through gas-grain chemistry. Although EA has been observed only in the gas phase and not within interstellar ices, it is hypothesized that its presence results from the erosion of icy mantles by low-velocity shocks in massive molecular clouds \citep{rivilla2021}. EA has also been found in the Almahata Sitta meteorite \citep{glavin2010}. In a recent study, we concluded that the detection of EA ice would be very challenging based on the estimation of an upper limit abundance in the solid state \citep{Suhasaria2024}.

EA, an aminoalcohol containing both amino (NH$_2$) and hydroxyl (OH) groups, is significant due to its potential role in the origin of life. As the simplest phospholipid head group, EA, also known as $\beta$-aminoethanol or glycinol, serves a precursor to glycine in prebiotic chemistry in simulated archean alkaline hydrothermal vents \citep{zhang2017}. It is also a component of various biomolecules, including aminosugars, sphingolipids, and glycoproteins, which are crucial for biological systems \citep{sladkova2014}.

Several formation pathways for EA in space have been proposed. One route involves the hydrogenation of aminoketene (NH$_2$-CH=C=O), which can form from CO and NH$_3$ ice with carbon atoms \citep{Krasnokutski2021, Krasnokutski2022}. Another route involves the ultraviolet irradiation of a mixture of H$_2$O, CH$_3$OH, NH$_3$, and HCN ices, producing EA along with other amino acids like glycine, alanine, and serine \citep{bernstein2002}. Additionally, the isomer of EA, $\alpha$-aminoethanol, can be synthesized through the thermal reaction of NH$_3$ with CH$_3$CHO, both in the presence and absence of H$_2$O \citep{duvernay2010}. In addition, only one very recent study has examined the destruction pathways of pure EA ice and EA mixed with water ice under ion and electron irradiation \citep{zhang2024}. However, this study only detected simple species following EA destruction, leaving open questions about their potential role as precursors to more chemically complex molecules.

In this study, we report the results of experiments where pure EA ice films were exposed to Ly-$\alpha$ (10.2 eV, 121.6 nm) radiation at 10 K. The resulting products from UV-induced destruction were analyzed using Fourier-transform infrared (FTIR) spectroscopy and quadrupole mass spectrometry, allowing us to determine the destruction cross-section of the parent EA molecule. By irradiating these low-temperature ices with UV photons and subsequently warming them to room temperature, it is possible to study the chemical processes that can occur in ices in star-forming regions near molecular clouds and protoplanetary disks. In these environments, more complex species are likely to be formed through radical-radical recombination. Therefore, the refractory residues left on the substrate at the end of experiments were further analyzed using FTIR spectroscopy. In the end, astrophysical implications regarding this work will be presented.

\section{Experimental methods}
\subsection{Experimental setup}

The experiments were conducted in the Laboratory Astrophysics group at Jena within the INterStellar Ice Dust Experiment (INSIDE) chamber. A detailed description of the setup is available in previous work \citep{potapov2019}, but a brief overview relevant to this study is provided here. The measurements took place in an ultrahigh vacuum (UHV) chamber, which maintained a base pressure below 8$\times$10$^{-10}$ mbar. This chamber is equipped with a closed-cycle helium cryostat that can cool samples down to 10 K. For our experiments, we used either a KBr or silicon window as the substrate, which was affixed to the cold finger of the cryostat for subsequent measurements.

Liquid NH$_2$CH$_2$CH$_2$OH (Sigma Aldrich, $\geq$99.0\%, EA-$^{12}$C$_2$) was degassed via multiple freeze-pump-thaw cycles and deposited as pure vapor onto the substrate using a gas inlet system connected to the UHV chamber. On the other hand, liquid NH$_2$$^{13}$CH$_2$$^{13}$CH$_2$OH (Sigma Aldrich, 99 atom \% $^{13}$C, EA-$^{13}$C$_2$) was available in a small vial with minute quantity so it was used without further purification. The chamber is equipped with a Fourier transform infrared (FTIR) spectrometer (Bruker Vertex 80v) and a Mercury Cadmium Telluride (MCT) detector for in situ ice phase measurements. Infrared spectra were collected during irradiation experiments at 10 K and during warming-up experiments across a temperature range of 10 to 300 K. These measurements were taken in transmittance mode, covering the spectral range from 6000 to 400 cm$^{-1}$ with a resolution of 1 cm$^{-1}$. For irradiation experiments at 10 K, 256 scans were recorded, while 32 scans were taken during each warming-up interval.

The effects of vacuum ultraviolet (VUV) irradiation on the grains were studied using a microwave-discharge H$_2$-flow lamp, with samples irradiated for approximately 3 hours at 10 K. The H$_2$ gas at a pressure of about 1-2 mbar (Air Liquide, 99.9\% purity) was excited by a 2450 MHz microwave generator (SAIREM GMP 03 KSM B) connected to an Evenson cavity (Opthos) providing 90 W forward power with a typical reflected power of 2 W. Based on the method described in previous work \citep{fulvio2014}, the Ly-$\alpha$ photon flux at the sample position was estimated to be (25 $\pm$ 9)$\times$ 10$^{12}$ photons cm$^{-2}$ s$^{-1}$. 


A quadrupole mass spectrometer (QMS; HXT300M, Hositrad) connected to the UHV chamber monitored the gas phase during the low-temperature and warming-up experiments to detect molecules desorbing from the sample surface. Temperature programmed desorption (TPD) experiments were performed by linearly ramping the sample temperature at a rate of 3 K min$^{-1}$ over a range of 10 to 300 K. The temperature measurement error was determined to be $\pm$ 1 K.

\subsection{Ice growth}

The deposition of EA-$^{13}$C$_2$ was done at 160 K to prevent and minimize the condensation of CO, CO$_2$ and H$_2$O on the substrate. During the deposition the QMS and IR signals were also monitored. This temperature is also below the phase transition of EA ice from amorphous to crystalline \citep{Suhasaria2024}. In previous work, we observed slight shifts in the fundamental vibrational modes of EA at this temperature, with redshifts in $\nu_\mathrm{as}$(NH$_2$), $\nu_\mathrm{as}$(CH$_2$), and $\tau$(NH, OH), and blue shifts in $\nu$(OH), $\nu_\mathrm{s}$(CH$_2$), and $\delta$(CH$_2$). No change in positions were detected for the remaining infrared bands. To further investigate the shifts in band position due to isotopic substitution, we compared the infrared spectra of the two EA isotopes at 160 K, and the results are summarized in Table \ref{table1}. Isotopic substitution caused mostly redshifts in bands involving two carbons or a carbon and a heteroatom (excluding hydrogen), but we also noted slight blue shifts in $\nu$(OH) and $\delta$(NH$_2$).

The thickness of the EA-$^{12}$C$_2$ ice film was estimated using the -NH$_2$ bending mode centered around 1607 cm$^{-1}$, based on a band strength value of $1\times10^{-17}$ cm molecule$^{-1}$ as determined in our recent work \citep{Suhasaria2024}. The average thickness was calculated to be 60 nm equivalent to about 60 ML. For this estimation, the density of EA ice was assumed to be the same as that of liquid EA at 20 $^{\circ}$C, which is 1.01 g cm$^{-3}$, due to the absence of direct measurements of the density of EA ice. The thickness of the EA-$^{13}$C$_2$ of about 60 ML was also estimated in the same way.

We evaluate the mean free path (MFP) of UV photons through the EA films and the fraction of energy absorbed by the films to get an insight on the impact of UV irradiation. To perform this evaluation, the UV absorption coefficient is required, as MFP is inversely proportional to this coefficient. The UV absorption coefficient is the product of the UV photo-absorption cross-section of EA and the number of EA molecules per unit volume. Since direct measurements of the Ly-$\alpha$ photo-absorption cross-section for EA ice are not available, we used the value determined for the Ly-$\alpha$ dissociation cross-section of ethanol and the ionization cross-section of methylamine, which is approximately $2.5\times10^{-17}$ cm$^2$ \citep{heays2017}. Based on this, the MFP was estimated to be around 40 nm. Then simply applying the concept of exponential attenuation, one can estimate the fraction of energy absorbed in the ice to be about 0.78. 

\startlongtable
\begin{deluxetable*}{cccc}
\tablecaption{Infrared absorption assignments for $^{12}$C EA ice deposited at 10 K or warmed to 160 K before irradiation, and for $^{13}$C EA ice deposited at 160 K \label{table1}}
\tablewidth{0pt}
\tablehead{
\colhead{Assignment} & \colhead{$^{12}$C at 10 K (cm$^{-1}$)} & \colhead{$^{12}$C warmed to 160 K (cm$^{-1}$)} & \colhead{$^{13}$C deposited at 160 K (cm$^{-1}$)} 
}
\startdata
$\nu_\mathrm{as}$(NH$_2$) & 3351 & 3338 & 3338\\   
$\nu_\mathrm{s}$(NH$_2$)& 3286 & 3286 &3286 \\
$\nu$(OH) & 3185 & 3200  & 3205\\
$\nu_\mathrm{as}$(CH$_2$) &2928  & 2925 &2916 \\ 
$\nu_\mathrm{s}$(CH$_2$) & 2857 & 2860 &2860 \\ 
$\delta$(NH$_2$) &1607  &1607  & 1615\\
$\delta$(CH$_2$) &1458  & 1470 &1470\\
$\nu$ (CC)+ $\delta$ (OH, CH) & 1395 &1395  &1380\\
$\tau$ (NH, OH) &1360  &1355  &1340\\
Skeletal vibrations &1253  & 1253 &1253\\
Skeletal vibrations &1175  &1175  &1171\\
$\nu$ (CN, CO) +$\delta$ (OH)+ $\tau$ (CC) & 1082 &1082  &1060\\
$\nu$ (CO)+ $\delta$ (NH, CH)+ $\tau$ (CC) &1033  & 1038 &1020\\
\hline
\enddata
\tablecomments{IR spectral assignments are based on \citep[and references therein]{Suhasaria2024}}
\end{deluxetable*}

\section{Results and discussion}
\subsection{Destruction cross-section}
\begin{figure}[!ht]
    \centering
    \resizebox{\hsize}{!}{\includegraphics{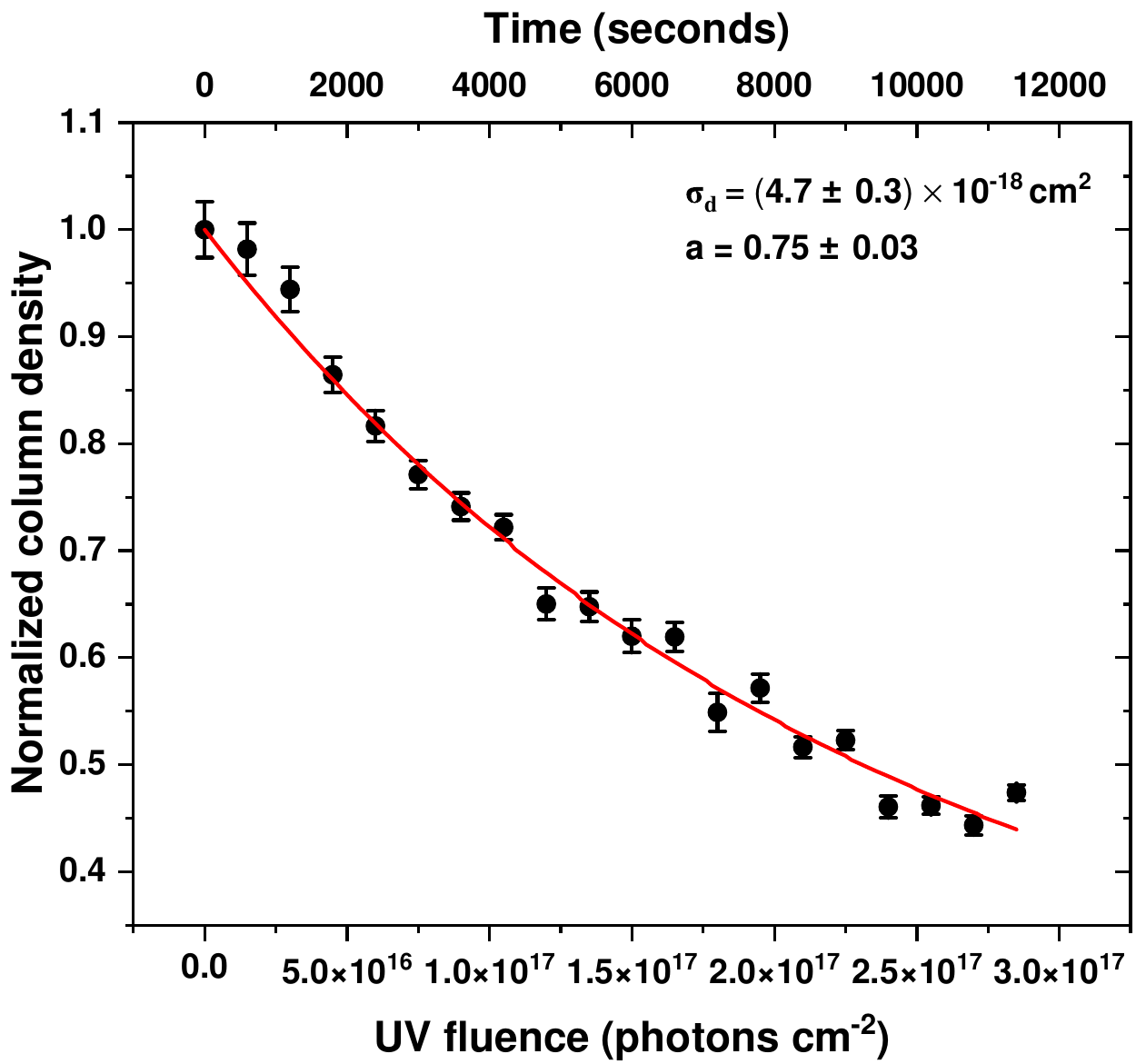}}
    \caption{Normalised column density of total C-H stretching band as represented by filled black circles for EA ice on KBr at 10 K, as the UV photon fluence increases. These values are normalised using the corresponding initial column densities. The error arising from baseline subtraction was considered in the error estimation. The best fit of Eq.~\ref{A} to the data is depicted with the solid red line. The estimated destruction cross-section ($\sigma_{d}$) and the asymptotic destruction ($\textit{a}$) are also provided.}
    \label{cross-section}
\end{figure}

Upon UV photolysis of pure EA-$^{12}$C$_2$ ice deposited at 10 K, the intensity of several infrared bands in EA decreases. Figure \ref{cross-section} illustrates the decrease of the normalized column density, $N(t)$ (unitless), of the $\nu$(CH$_2$) bands (combined asymmetric and symmetric modes) in EA-$^{12}$C$_2$ ice as UV fluence, $F_{UV}$, increases. The experimental data is fitted using the first-order kinetics with the following equation:
\begin{equation}
\label{A}
\mathrm{N(t)} = \mathrm{1- a(1- e^{-\sigma_{d}F_{UV}})}
.\end{equation}

The equation deals with the destruction process under the boundary condition that incorporates a residual reactant at $\mathrm{F_{UV}}$ = $\infty$. $N(t)$ represents the number of molecules remaining  at a time $t$ and $\textit{a}$ (asymptotic value) is the fraction of molecules destroyed when exposed to UV radiation. Our analysis of the experimental data allowed us to deduce the cross-section ($\mathrm{\sigma_{d}}$) and $\textit{a}$ for EA destruction. It is important to note that while we used the $\nu$(CH$_2$) bands to calculate the destruction cross-section, new species formed after irradiation could also contribute to the increase in this band. However, as we observed a consistent decrease in the signal intensity with increasing UV fluence, any such effect on the cross-section evaluation would be minimal. Therefore, the cross-section estimation using this band would be fairly appropriate.

After nearly 180 minutes of Ly-$\alpha$ irradiation, about 55\% of the parent species is consumed. The photo destruction cross-section is estimated to be ($4.7\pm0.3)\times10^{-18}$ cm$^2$. In a similar experiment, the photo destruction cross-section of $\alpha$-aminoethanol, an isomer of EA, for VUV irradiation ($\lambda$ \textgreater 120 nm) was found to be an order of magnitude lower \citep{duvernay2010}. The possible deviation could arise due to the difference in stability of $\alpha$-aminoethanol with respect to EA. In a recent study \citep{zhang2024}, the destruction cross-section of pure EA ice film at 20 K by 1 keV electron bombardment was estimated as $2\times10^{-16}$ cm$^2$ which is nearly 50 times larger than our measurement. This result can be understood because in electron-induced interactions a large number of low-energy secondary electrons also interact with a solid. In contrast, Ly-$\alpha$ photons typically induce a single photolysis event. Therefore, the UV destruction cross-section should be lower than that for electrons.

\subsection{New products formation after UV irradiation}

\begin{figure}[!ht]
    \centering
     \resizebox{\hsize}{!}{\includegraphics{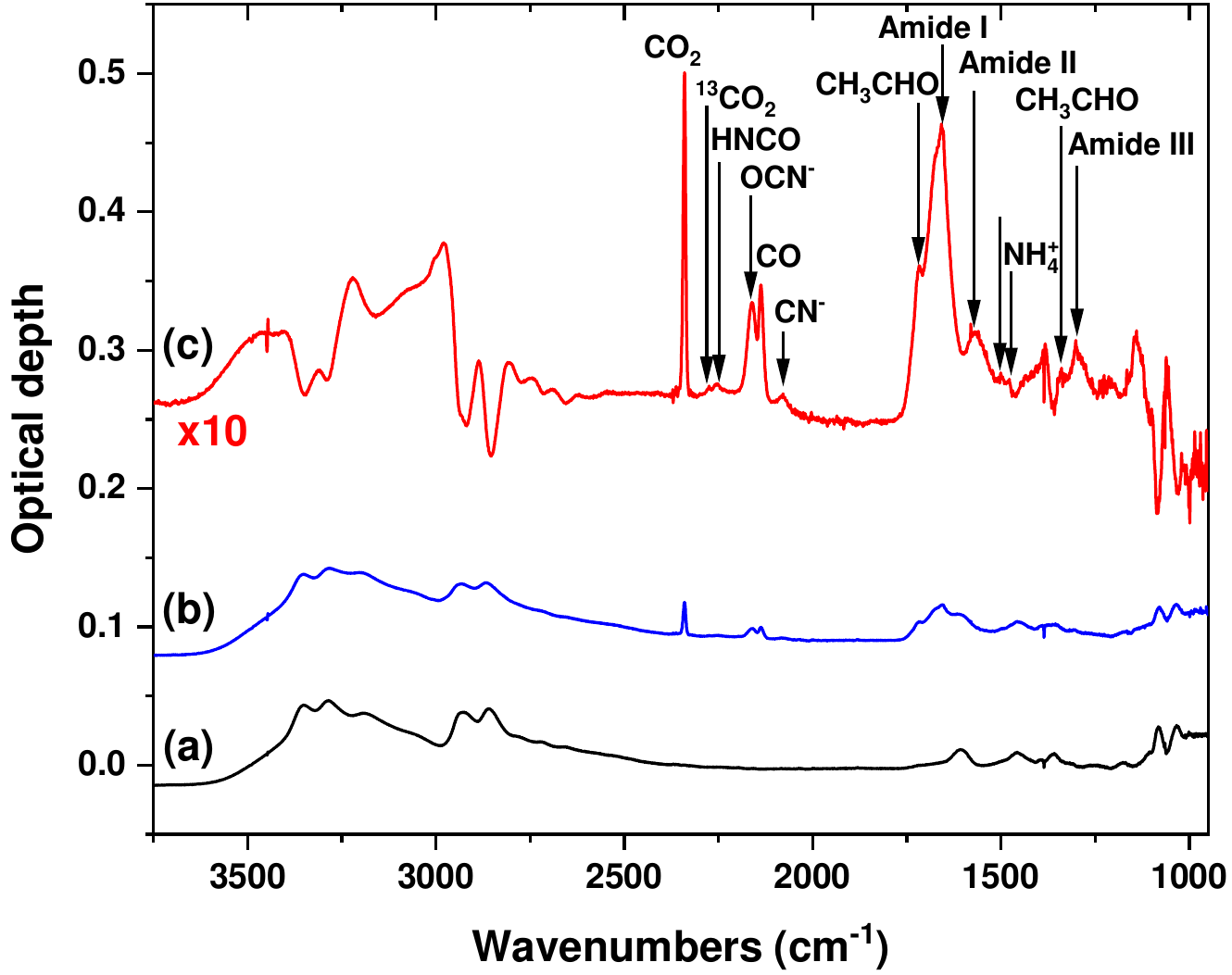}}
\caption{Mid-IR spectrum of about 60 ML EA-$^{12}$C$_2$ film on KBr at 10 K, shown as deposited (a, black line) and after UV irradiation with 2.8$\times10^{17}$ photons cm$^{-2}$ (b, blue line). The new products following irradiation are evident in the difference spectrum (c, red line), which represents the spectrum after irradiation minus the spectrum before irradiation. Negative bands correspond to the parent species, while positive bands indicate new products. For clarity, the blue and red curves are offset along the ordinate axis.}
    \label{products}
\end{figure}

We observe not only the destruction of the parent species but also the formation of new compounds, as indicated by the appearance of new infrared bands as seen in Figure \ref{products}. The most intense of these bands were readily identified as the stretching vibrations of carbon dioxide (CO$_2$), cyanate ion (OCN$^{-}$), and carbon monoxide (CO). In the 2500–2000 cm$^{-1}$ region, weak bands appear at 2276, 2255, and 2080 cm$^{-1}$. The band at 2276 cm$^{-1}$ likely corresponds to the stretching vibration of $^{13}$CO$_2$, while the band at 2255 cm$^{-1}$ can be attributed to the N=C=O functional group, with the asymmetric stretching mode of the isocyanic acid (HNCO) being a strong candidate. The 2080 cm$^{-1}$ band is likely due to hydrogen cyanide (HCN), cyanide ion (CN$^{-}$), or a combination of both.

The counter ion ammonium (NH$_4^+$) for OCN$^{-}$ and CN$^{-}$ was observed from the NH IR bending mode feature around 1480 cm$^{-1}$ \citep{van2004}. In the thermal reaction of NH$_3$ and HCN, NH$_4^+$ infrared features vary between 1470 and 1500 cm$^{-1}$ depending on the ice composition in which it is present \citep{noble2013}.

In the 1800--1200 cm$^{-1}$ region, a strong and broad absorption band with a double peak structure centered around 1675 and 1655 cm$^{-1}$ appears, along with two weaker broad bands peaking around 1573 and 1300 cm$^{-1}$. These features are consistent with typical amide (R-CONH$_2$) I (C=O stretching), II (out-of-phase combination of the NH in-plane bend and the CN stretching vibration), and III (in-phase combination of the NH bending with CN stretching and NH deformation modes) bands, as observed in previous experiments involving UV photoirradiation of methylisocyanate or glycine at low temperatures \citep{mate2018}. However, there are alternative interpretations of these bands. For example, the amide I band could be attributed to C=O stretching in ketones or carboxylic acids, or C=C stretching in alkenes. The amide II band could correspond to asymmetric C=O stretching in carboxylic acids, while the amide III band could be related to C-O stretching and different types of CH bending modes \citep[e.g.][]{kim2010, james2021}.

In the 1800–1000 cm$^{-1}$ region, several weak peaks are observed. The peak at 1720 cm$^{-1}$ is typically attributed to C=O stretching in carbonyl-containing species such as aldehydes, ketones and carboxylic acids. Based on our infrared data, we identified features consistent with both acetaldehyde (CH$_3$CHO) and formaldehyde (H$_2$CO). For CH$_3$CHO, we identified two additional peaks at 1434 cm$^{-1}$ and 1341 cm$^{-1}$, corresponding to CH$_3$ deformations \citep{hudson2020}. Formaldehyde exhibits a weak band at 1497 cm$^{-1}$ \citep{gerakines1996}. If the amino radical is formed and long-lived, it would be detectable via its bending mode at 1499 cm$^{-1}$, which may overlap with the H$_2$CO feature \citep{marks2023}. The peak at 1060 cm$^{-1}$ was assigned to ethanol (C$_2$H$_5$OH) \citep{van2018}. 

The production of HCN, HNCO, and OCN$^{-}$ after UV photoprocessing of EA ice in this experiment is reminiscent of the reactions involving formamide \citep{suhasaria2022}. Thus, these species could also be secondary photoproducts, formed after amides are produced as first-generation products in the ice mixture. However, given the complexity of the reaction products, it is unclear which specific amide is initially formed.

Although the species identified from the destruction of EA by 1 keV electrons \citep{zhang2024} are somewhat similar to those in our study, there are notable differences. We cannot clearly assign an infrared feature to NH$_3$ ice, as its N-H stretching (between 3100 and 3500 cm$^{-1}$) and umbrella mode (around 1100 cm$^{-1}$) are heavily masked by the presence of the parent EA ice, while the N-H bending mode (around 1630 cm$^{-1}$) is obscured by the formation of a broad amide band. Additionally, we find no evidence of CH$_3$OH (which appears around 1030 cm$^{-1}$ \citep{caro2003}) or C$_2$H$_4$. While we observe an infrared band near 1440 cm$^{-1}$, close to the CH$_2$ scissoring region, we do not detect the strongest infrared feature of pure C$_2$H$_4$ ice at 950 cm$^{-1}$ (due to CH$_2$ wagging) \citep{hudson2014, zhou2014}. Therefore, the 1440 cm$^{-1}$ feature is more confidently assigned to acetaldehyde \citep{van2018}. We also cannot confidently assign infrared features to H$_2$O or H$_2$O$_2$.

To ensure a clear identification of the products, we have compared infrared results of $^{12}$C with $^{13}$C-labeled EA. We noticed a red-shift in the band positions for $^{13}$C as expected. These shifts, along with the assigned products and their corresponding wavenumbers, are summarized in Table \ref{table2}.

\startlongtable
\begin{deluxetable*}{ccccc}
\tablecaption{Infrared absorption assignments for the products formed after irradiation of the two EA isotopes\label{table2}}
\tablewidth{0pt}
\tablehead{
\colhead{Species} & \colhead{Assignment} & \colhead{$^{12}C_2$ EA (cm$^{-1}$)} & \colhead{$^{13}$$C_2$ EA (cm$^{-1}$)} & \colhead{Reference} 
}
\startdata
CO$_2$ & $\nu_\mathrm{a}$(CO) & 2340 & 2340 & 1 \\
$^{13}$CO$_2$ & $\nu_\mathrm{a}$(CO) &2274 & 2274 & 2\\
HNCO & $\nu$(NCO) & 2255 & 2255 & 3\\
HN$^{13}$CO & $\nu$(NCO) &  & 2205 & \\
OCN$^-$ & $\nu$(C$\equiv$N) & 2160 & & 4\\
O$^{13}$CN$^-$ & $\nu$(C$\equiv$N) & & 2107 & 5\\
CO & $\nu$(CO) &2135  & 2135& 6 \\
$^{13}$CO & $\nu$(CO) &  & 2090& 2 \\
HCN and CN$^-$ & $\nu$(C$\equiv$N)&2078  & & 7\\
H$^{13}$CN and $^{13}$CN$^-$ & $\nu$(C$\equiv$N)&  & 2040& 5\\
CH$_3$CHO & $\nu$(C=O) & 1720 & & 8\\
$^{13}$CH$_3$$^{13}$CHO & $\nu$(C=O) &  &1680 & 8\\
H$_2$CO & $\nu$(C=O) & 1720 & & 9\\
H$_2^{13}$CO & $\nu$(C=O) &  & 1680& \\
Amide I& $\nu$(CO) & 1657 &  & 10, 11\\
$^{13}$C-Amide I& $\nu$(CO) &  & 1618 & 11\\
Amide II & out-of-phase $\nu$(CN)+$\delta$(NH) & 1567 &  & 10, 11\\
$^{13}$C-Amide II& out-of-phase $\nu$(CN)+$\delta$(NH) &  &1555  & \\
NH$_2$CH$_2$CHO & $\nu$(C=O) & 1572 & 1566 & 12\\
H$_2$CO & $\delta$(CH) & 1497 & &9\\
NH$_4^+$& $\delta$(NH) & 1478 & 1478 & 7\\
CH$_3$CHO & $\delta${CH$_3$} & 1434 &  & 8\\
$^{13}$CH$_3$$^{13}$CHO & $\delta${CH$_3$} &  & 1424 & 8\\
CH$_3$CHO & $\delta${CH$_3$} & 1341 &  & 8\\
$^{13}$CH$_3$$^{13}$CHO & $\delta${CH$_3$} &  & 1325 & 8\\
Amide III& in-phase $\nu$(CN)+$\delta$(NH) & 1301 & & 10\\
$^{13}$C-Amide III& in-phase $\nu$(CN)+$\delta$(NH) &  &1294 & \\
C$_2$H$_5$OH & $\nu$(CO) & 1060 &  & 13\\
$^{13}$C$_2$H$_5$OH & $\nu$(CO) &  & 1040 & \\
\hline 
\enddata
\tablerefs{\citet{yamada1964}[1], \citet{ehrenfreund1997}[2], \citet{lowenthal2002}[3], \citet{grim1989}[4], 
\citet{pozun2012}[5], \citet{hagen1980}[6], \citet{noble2013}[7], \citet{hudson2020}[8], \citet{gerakines1996}[9], \citet{mate2018}[10], \citet{mukherjee2004}[11], \citet{marks2023}[12]
\citet{van2018}[13]}
\end{deluxetable*}

\subsection{Mass and infrared spectra during warming up}

\begin{figure}[!ht]
    \centering
    \resizebox{\hsize}{!}{\includegraphics{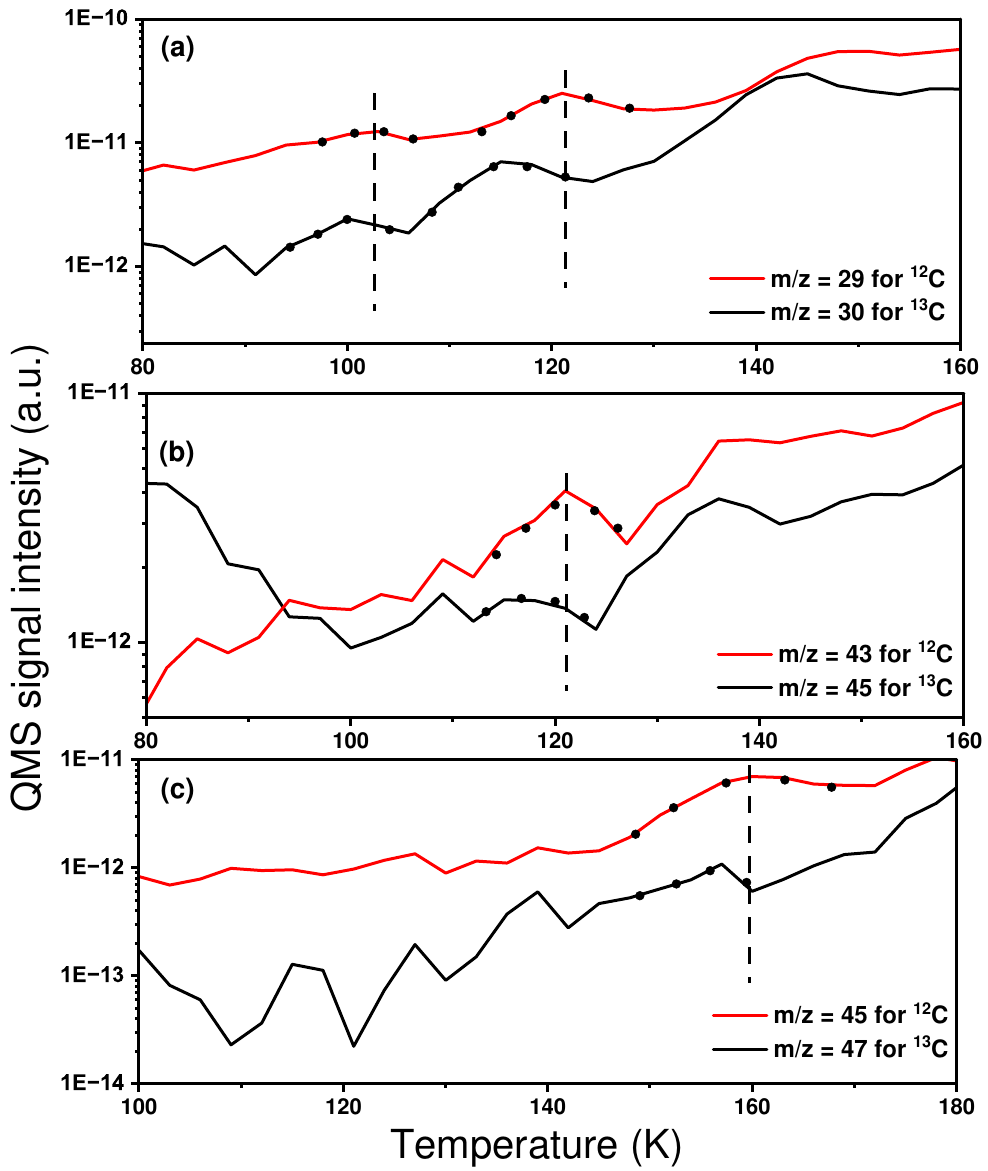}}
\caption{TPD traces obtained via QMS following the conclusion of UV irradiation. The mass signals at m/z = 29 (HCO) and 30 (a), m/z= 43 (H$_3$C$_2$O), and 45 (b), and m/z= 45 (H$_5$C$_2$O), and 47, correspond to the $^{12}$C and $^{13}$C isotopes of the products.} The signals could be associated to the desorption of H$_2$CO, CH$_3$CHO, and C$_2$H$_5$OH, respectively. The dotted lines on the TPD traces are guide to the eye.
    \label{TPD}
\end{figure}

\begin{figure}[!ht]
    \centering
    \resizebox{\hsize}{!}{\includegraphics{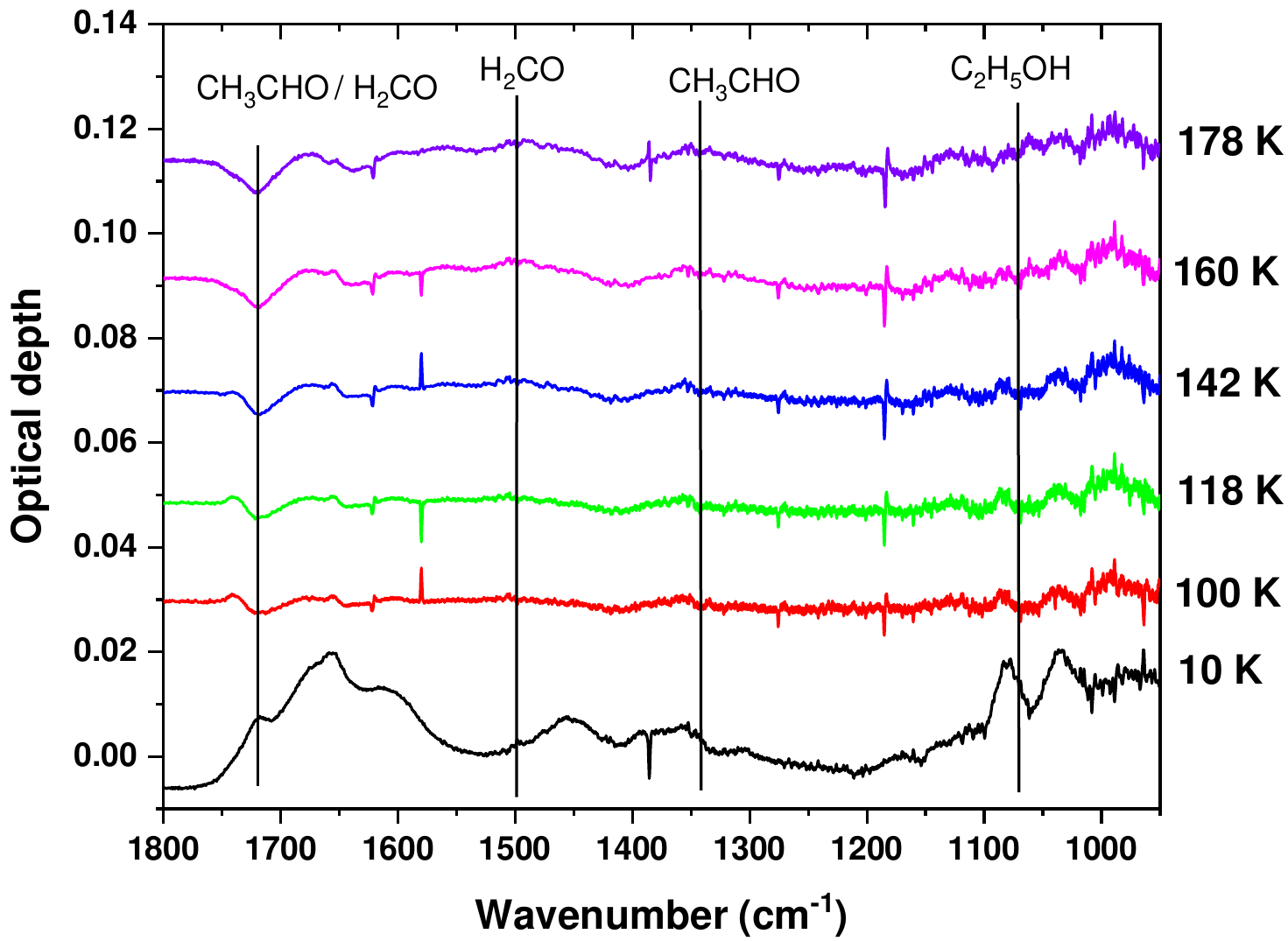}}
\caption{Mid-IR spectra of the ice after UV irradiation and during TPD. The vertical lines indicate the infrared band positions of H$_2$CO, CH$_3$CHO, and C$_2$H$_5$OH. For clarity, the curves depicting the TPD spectra are offset along the ordinate axis}.
    \label{IR}
\end{figure}

Identifying products based only on infrared features is not conclusive. Thus, we also examined mass peaks using QMS during TPD and monitored the corresponding infrared signals during warming up. We observed desorption of H$_2$CO, CH$_3$CHO, and C$_2$H$_5$OH at around 100 K, 120 K, and 160 K, as shown in Figure \ref{TPD} (a), (b) and (c), respectively, for $^{12}$C. Corresponding signals for $^{13}$C are also shown in the same Figure. The reference mass spectrum for the $^{12}$C products was taken from the National Institute of Standards and Technology (NIST) database. First, we observed that the desorption signals arising from $^{13}$C appeared at about 5 K lower than those for $^{12}$C for all the products. Although isotope substitution typically does not affect binding energy, this shift may be due to the different morphology of the starting EA-$^{13}$C$_2$ before irradiation, possibly resulting from a higher deposition temperature (160 K). Second, only a small amount of molecules desorbed at the typical desorption temperatures for the pure species. The majority of the molecules remained trapped in the matrix of unreacted EA (about 25\%) and desorbed later, during the phase transition of EA (around 180 K) as seen in the Figure \ref{IR}. Some of the products may also co-desorb at higher temperatures during EA desorption. Another possibility is that the molecules undergo further reactions during warming up which could account for the low desorption signals.

\begin{figure*}[!ht]
\plottwo{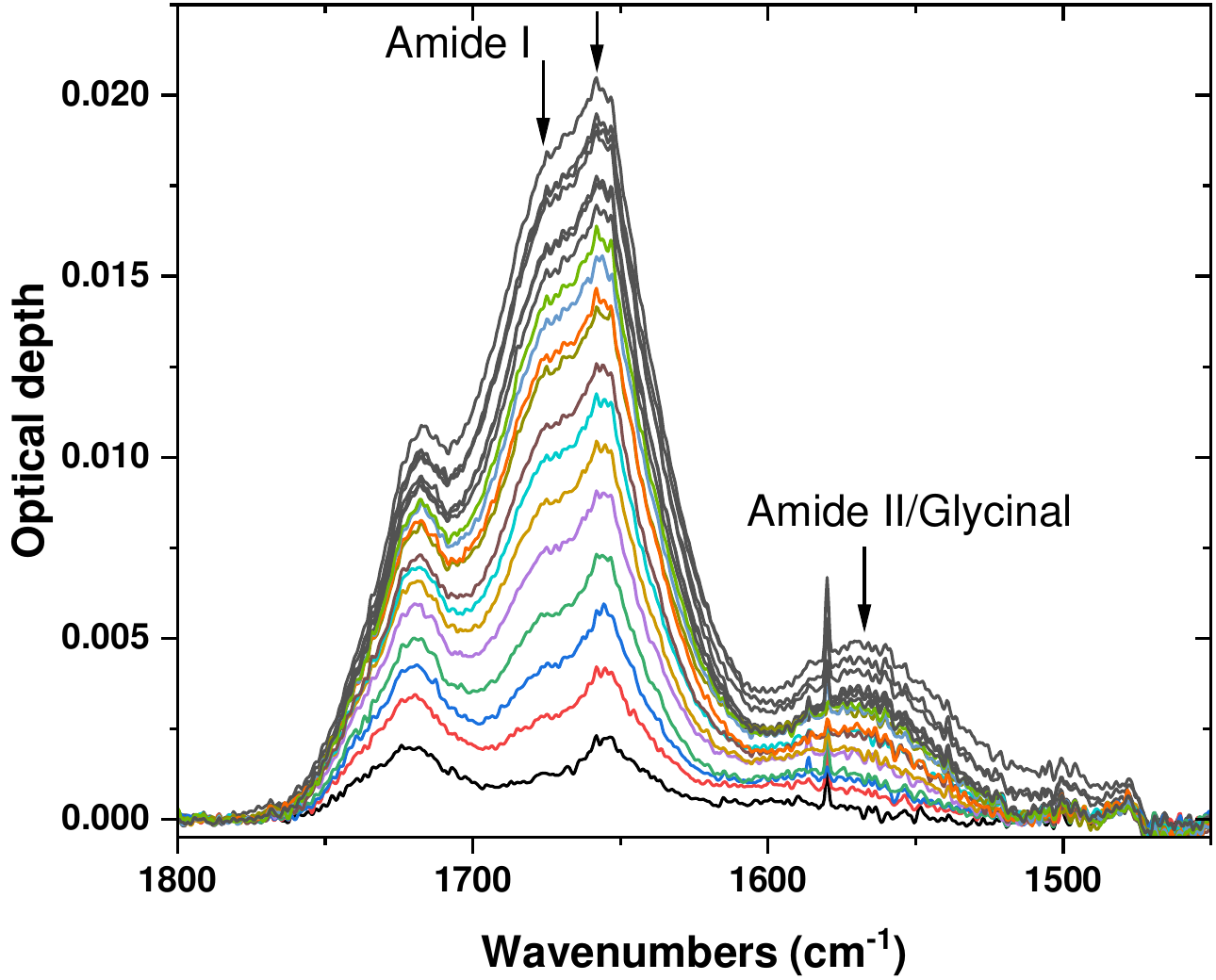}{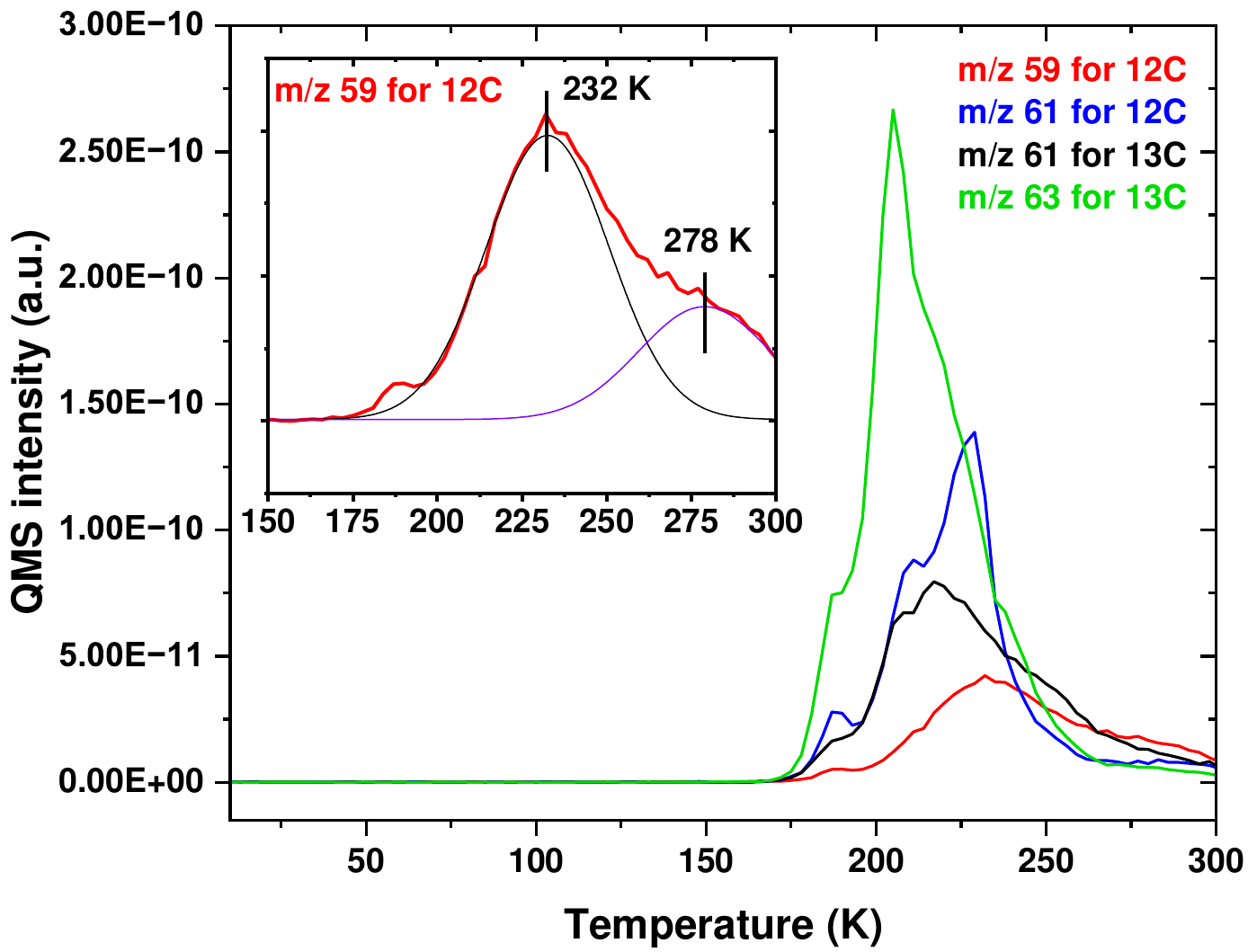}
\caption{The left-hand panel shows a series of mid-IR spectra, where the increasing UV fluence clearly leads to the growth of the amide I and II peaks, along with glycinal in EA-$^{12}$C$2$. In the right-hand panel, the TPD traces obtained via QMS after UV irradiation are displayed. The mass signals at m/z= 59 and 61 correspond to the desorption of glycinal tautomer and EA-$^{12}$C$_2$, respectively. The inset highlights the mass signal at m/z= 59, with peaks around T = 232 K and 278 K, likely corresponding to the desorption of 2-aminoethenol and 1-aminoethenol, respectively. For comparison, the mass signals at m/z= 61 and 63 corresponding to glycinal tautomer and EA-$^{13}$C$_2$ are also shown.  \label{glycinal}}
\end{figure*}

\begin{figure}
    \centering
    \resizebox{\hsize}{!}{\includegraphics{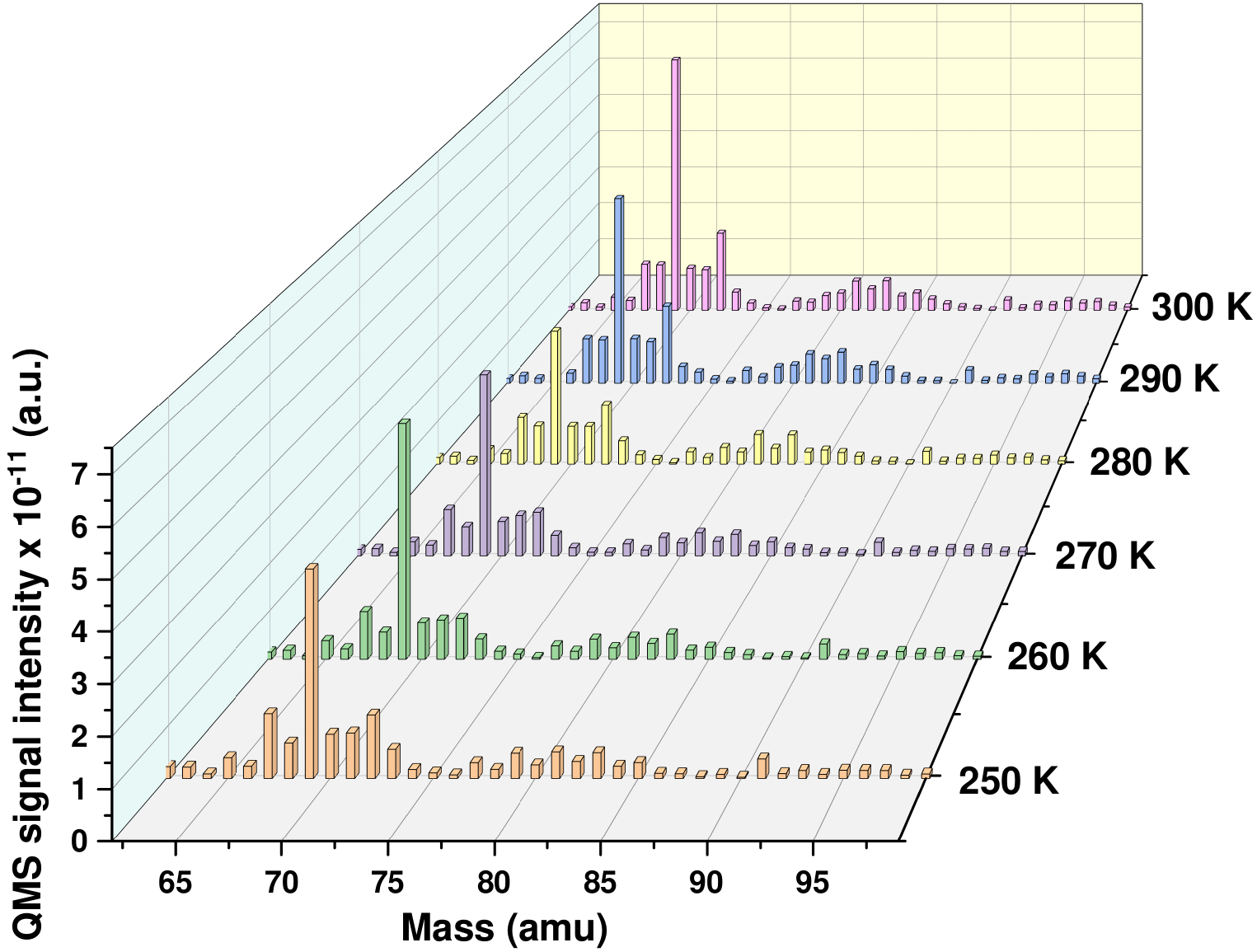}}
    \caption{Mass spectra during TPD after UV irradiation of EA-$^{12}$C$_2$ are shown in the m/z range between 62 and 99 at different temperatures starting from 250 K. The spectra reveal the appearance of several high mass peaks that cannot originate from the parent species, suggesting the formation of complex organic molecules that even remains on the surface at room temperature.}
    \label{COMs}
\end{figure}

\begin{figure}
    \centering
    \resizebox{\hsize}{!}{\includegraphics{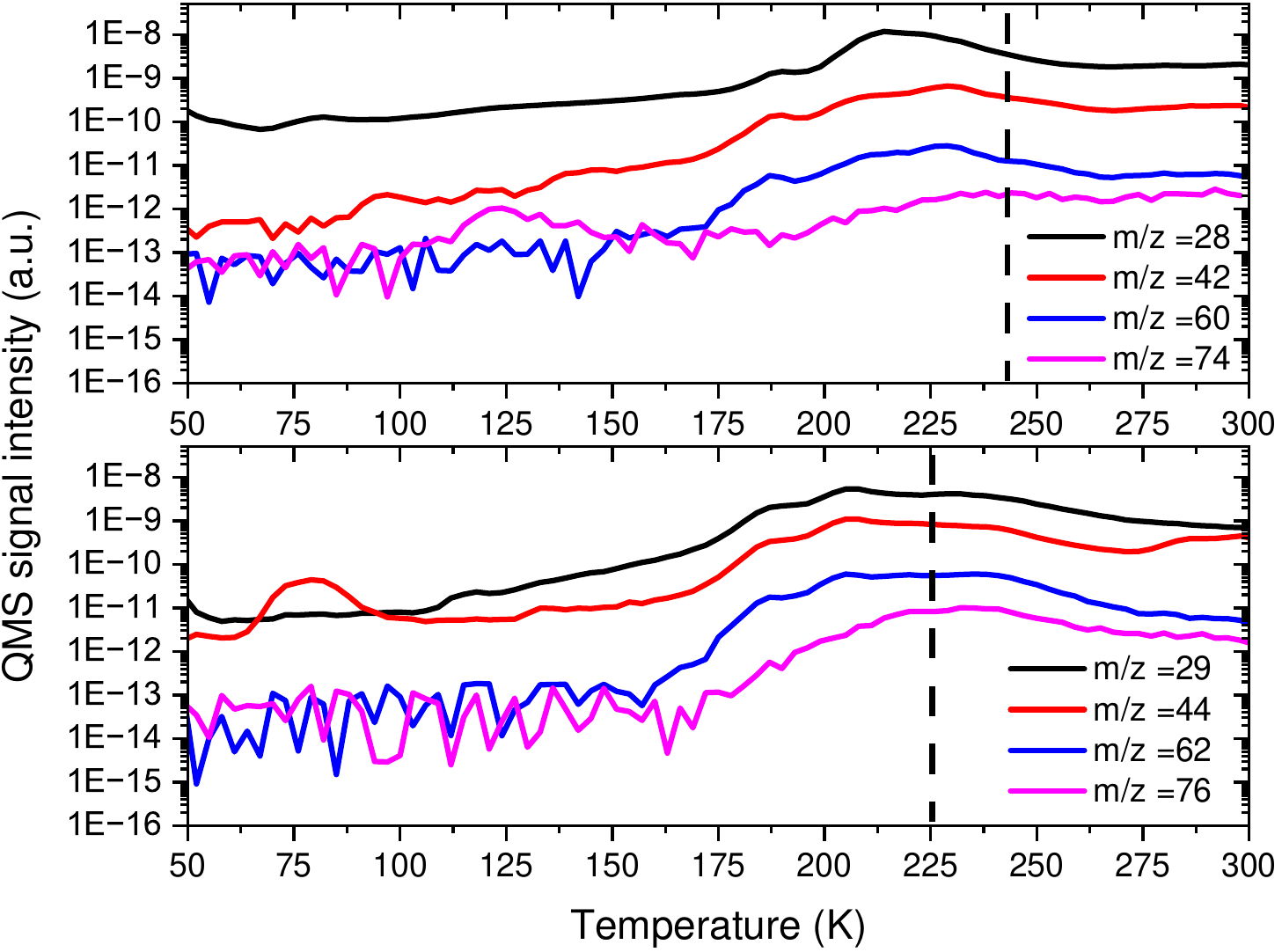}}
    \caption{Ethylene glycol is identified from the TPD traces after UV irradiation of EA ice. The mass signals at m/z= 31 (CH$_2$OH$^+$), 33 (CH$_3$OH$_2$$^+$), and 62 (CH$_2$OH)$_2$$^+$ in the top panel and m/z= 32, 34 and 64 in the bottom panel, corresponding to the $^{12}$C and $^{13}$C isotope of EG, respectively. The mass spectrum of EG was taken from the NIST database.}
    \label{EG}
\end{figure}

\begin{figure}
    \centering
    \resizebox{\hsize}{!}{\includegraphics{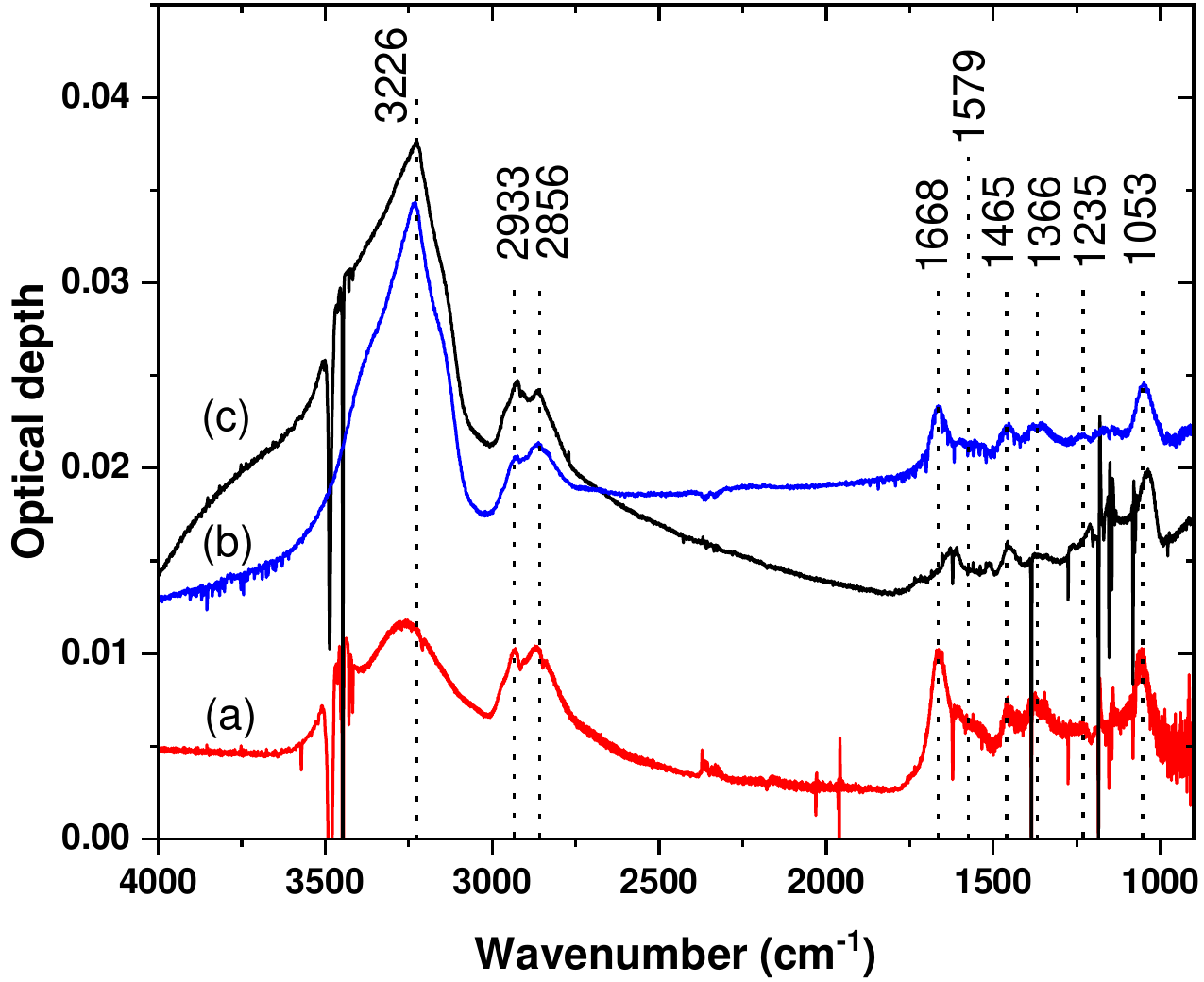}}
    \caption{Serine is tentatively identified from the TPD traces after UV irradiation of EA ice. The mass signals at m/z= 28 (NHCH$^+$), 42 (NH$_2$)C$_2$H$_2$$^+$), 60 (OHCH$_2$CH(NH$_2$)$^+$), and 74 (COOH(NH$_2$)CH$^+$)} in the top panel and m/z=  29, 44, 62, and 76 in the bottom panel, corresponding to the $^{12}$C and $^{13}$C isotope of Serine, respectively. The mass spectrum of Serine was taken from a recent study \citep{farajmand2016}.
    \label{Ser}
\end{figure}

The band observed at 1573 cm$^{-1}$ can also be attributed to the C=O stretching vibration of the formyl methyl, also called vinoxy radical, ($\cdot$CH$_2$CHO) in addition to the Amide II bands as seen in the left-hand panel of Figure \ref{glycinal}. Previously, this feature has been reported in two contexts. First, it appeared when acetaldehyde ice was irradiated with 1 MeV protons at 20 K \citep{hudson2018}, and second it was associated with glycinal (NH$_2$CH$_2$CHO) formed in interstellar ices containing acetaldehyde and ammonia following the impact of 5 keV electrons \citep{marks2023}. We attribute the 1573 cm$^{-1}$ peak to glycinal, which can form through the dehydrogenation of EA, similarly to how $\alpha$-aminoethanol (an isomer of EA) produces acetamide (CH$_3$CONH$_2$) under VUV radiation ($\lambda$ \textgreater 120 nm). Furthermore, glycinal is thermodynamically less stable and undergoes keto-enol tautomerism at higher temperatures, converting into 2-aminoethenol (HOCHCHNH$_2$), which desorbs at 232 K \citep{marks2023}. There may also be a formation of a small amount of 1-aminoethenol (isomer of 2-aminoethenol), which then desorbs at 278 K. This behavior can be observed in the TPD trace shown in the right-hand panel of Figure \ref{glycinal}. In the inset of Figure \ref{glycinal}, the possible TPD trace for glycinal has been fitted with two Gaussians, indicating the two desorption temperatures. To differentiate the TPD signal of m/z= 59 (associated to glycinal) from the mass fragments of EA, we also plot m/z= 61 for comparison. This shows that EA desorbs at a slightly lower temperature. The mass signals arising due to $^{13}$C isotopic substitution is also shown in the same Figure. Again, a shift of about 15 K to a lower temperature is observed in the desorption signals for $^{13}$C glycinal compared to $^{12}$C, which is similar to what we observed in the desorption temperatures for the two EA isotopes.
During the TPD several low-intensity peaks were observed, particularly in the mass range that could not have been produced by the desorption of the parent EA as shown in Figure \ref{COMs}. These signals were observed at higher masses in a temperature range from 250 K to room temperature. This observation suggests that some of the complex organic molecules may have desorbed before room temperature, while some others may have remained on the surface at room temperature as organic residues, which will be discussed later. Complete identification of all the complex organic molecules would be very challenging, but some of the tentative identifications are discussed below.
The mass signals at m/z= 62 together with m/z=31 and 33 showed a desorption peak around 230 K after UV irradiation of $^{12}$C EA, while the $^{13}$C counterpart showed a peak around 210 K as seen in the Figure \ref{EG}. These m/z signals can be associated with CH$_2$OH$^+$ (m/z=31, main fragment) and CH$_3$OH$_2$$^+$ (m/z=33) and (CH$_2$OH)$_2$$^+$ (m/z=62) fragments from Ethylene glycol (EG, (CH$_2$OH)$_2$) which can be potentially formed from the recombination of two $\cdot{\mathrm{C}}$H$_2$OH radicals forming from EA. Furthermore, the desorption temperatures of $^{12}$C species are consistent with previous TPD studies \citep{leroux2021}. The formation of EG could potentially be important since they are the simplest sugar alcohols that could form sugar or sugar derivatives.
We also searched for amino acids, but no clear signals for glycine or alanine were detected. We then examined serine (HOCH$_2$CH(NH$_2$)COOH), which has similarities in structure to EA, as serine contains the intact EA fragment. The mass signals at m/z = 60 (along with corresponding signals at m/z = 28, 42, and 74) are plotted as a function of temperature. A recent study on glycine formation in ice indicated that glycine desorbs at around 245 K \citep{ioppolo2021}. In the absence of other desorption studies on serine, it is reasonable to suggest that serine may also desorb at around this temperature or higher. Figure \ref{Ser} displays a desorption signal near this temperature following UV irradiation of $^{12}$C EA, though this is not the maximum desorption signal. It can be argued that these masses are not exclusive to serine, as other complex species might also desorb at this temperature and contribute to the signal. For comparison, the $^{13}$C counterpart exhibits a corresponding signal at 225 K. These signals could correspond to fragments of serine, including NHCH$^+$ (m/z= 28), (NH$_2$)C$_2$H$_2$$^+$ (m/z= 42), OHCH$_2$CH(NH$_2$)$^+$ (m/z= 60, main fragment, loss of -COOH group) and COOH(NH$_2$)CH$^+$ (m/z= 74). Serine could potentially be formed by the recombination of HOCH$_2$$\cdot{\mathrm{C}}$H(NH$_2$) radical with $\cdot{\mathrm{C}}$OOH formed from EA. Once amino acids are formed, they could potentially lead to the formation of peptides and peptide derivatives.
\subsection{Formation pathways for new product formation}

We can now discuss the formation pathways of the photoproducts identified in the previous sections. The formation of glycinal can be explained by a dehydrogenation mechanism, which has been previously observed during the VUV photolysis of aminomethanol and $\alpha$-aminoethanol \citep{bossa2009, duvernay2010}

\begin{equation}
\label{reaction1}
\mathrm{NH_2CH_2CH_2OH} +{\mathrm{h\nu}} \rightarrow \mathrm{NH_2CH_2CHO} + \mathrm{H_2}
.\end{equation}

The formation of ethanol and acetaldehyde requires the cleavage of the C-N bond, resulting in ammonia as a byproduct. \citet{sladkova2014} observed acetaldehyde and ammonia after UV photolysis of EA in aqueous solution using a mercury arc lamp ($\lambda$ 230–320 nm). Additionally, \citet{shibata2010} suggested that acetaldehyde and ammonia could be formed via initial H-atom abstraction from EA.

\begin{subequations}
\begin{equation}
\label{reaction2}
\mathrm{NH_2CH_2CH_2OH} +{\mathrm{h\nu}} \rightarrow \mathrm{\cdot{CH_2CH_2OH}} + \mathrm{\cdot{NH_2}}
,\end{equation}
\begin{equation}
\label{reaction3}
\mathrm{NH_2CH_2CH_2OH} +{\mathrm{h\nu}} \rightarrow \mathrm{\cdot{CH_2CHO}} + \mathrm{\cdot{NH_2}}+ \mathrm{H_2}
,\end{equation}
\begin{equation}
\label{reaction4}
\mathrm{\cdot{CH_2CH_2OH}} +{\mathrm{\cdot{H}}} \rightarrow \mathrm{CH_3CH_2OH}
,\end{equation}
\begin{equation}
\label{reaction5}
\mathrm{\cdot{CH_2CHO}} +{\mathrm{\cdot{H}}} \rightarrow \mathrm{CH_3CHO}
,\end{equation}
\begin{equation}
\label{reaction6}
\mathrm{\cdot{NH_2}} + {\mathrm{\cdot{H}}}\rightarrow \mathrm{NH_3} 
.\end{equation}
\end{subequations}

Once acetaldehyde is formed, it can dissociate to yield CO \citep{sampson2010}. Glycinal, being thermodynamically unstable, can also dissociate further, producing CO through an alternative pathway.

\begin{subequations}
\begin{equation}
\label{reaction7}
\mathrm{CH_3CHO} +{\mathrm{h\nu}} \rightarrow \mathrm{\cdot{CH_3}} + \mathrm{\cdot{CHO}}
,\end{equation}

\begin{equation}
\label{reaction8}
\mathrm{NH_2CH_2CHO} +{\mathrm{h\nu}} \rightarrow \mathrm{\cdot{NH_2}}+ \mathrm{\cdot{CH_2}} + \mathrm{\cdot{CHO}}
,\end{equation}

\begin{equation}
\label{reaction9}
\mathrm{\cdot{CHO}} + \mathrm{\cdot{H}} \rightarrow \mathrm{H_2} +\mathrm{CO} 
.\end{equation}

\end{subequations}

CO can drive the production of CO$_2$ \citep{molpeceres2023} and HNCO \citep{hudson2001} in the ice mixture through the following reactions:

\begin{subequations}
\begin{equation}
\label{reaction10}
\mathrm{CO} + \mathrm{\cdot{OH}} \rightarrow \mathrm{HOCO}\rightarrow\mathrm{\cdot{H}} + \mathrm{CO_2}
,\end{equation}
\begin{equation}
\label{reaction11}
\mathrm{CO} + \mathrm{\cdot{NH_2}} \rightarrow \mathrm{\cdot{H}} + \mathrm{HNCO}
, or \end{equation}
\begin{equation}
\label{reaction12}
\mathrm{CO} + \mathrm{\cdot{NH}} \rightarrow \mathrm{HNCO}
,\end{equation}
\end{subequations}

Once the HOCO intermediate is formed, it can react further with an R-group (where, R = H or alkyl) to form carboxylic acids \citep{kim2010}. 

\begin{equation}
\label{reactionX}
\mathrm{HOCO} + \mathrm{\cdot{R}} \rightarrow \mathrm{RCOOH} 
\end{equation}

CO could also react with subsequent H radicals to form H$_2$CO. The reaction between HNCO and NH$_3$ (via an acid-base reaction) produces ammonium cyanate salt \citep{van2004}:

\begin{equation}
\label{reaction13}
\mathrm{NH_3} + \mathrm{HNCO} \rightarrow \mathrm{NH_4^{+}OCN^{-}}
,\end{equation}

The CN$^{-}$ feature could result from the reaction between HCN and NH$_3$ \citep{noble2013}, where HCN might form as a secondary product from the processing of amides in the ice mixture, reminiscent of the UV photoprocessing of pure formamide \citep{suhasaria2022}. As already mentioned above, formamide could be formed from the recombination of NH$_2$ and HCO radicals since the process is barrierless \citet{chuang2022}. N-methyl formamide could also be formed in the ice mixture which could also contribute to the desorption peak shown in the right-hand panel of Figure \ref{glycinal}. 

\begin{equation}
\label{reaction14}
\mathrm{NH_3} + \mathrm{HCN} \rightarrow \mathrm{NH_4^{+}CN^{-}}
,\end{equation}

Besides the product pathways mentioned above, UV irradiation of EA ice can also produce small reactive radicals like OH, CH$_2$, CH$_2$OH, NH$_2$, and CH$_2$NH$_2$, which can add to the complexity of the products.

\subsection{Infrared spectrum of residues at room temperature}

The infrared spectrum of residues on the substrate after warming up UV-irradiated EA ice to 300 K shows several new peaks, indicating the emergence of functional groups that were not present in the irradiated ice. This suggests that additional species form during the warming process, driven by reactions of radicals formed by irradiation. As the temperature increases, radicals become more mobile and can meet each other. Additionally, new reactions with small energy barriers are allowed. All this leads to the formation of new chemical substances. Simultaneously, volatile compounds desorb, leaving behind stable, non-volatile molecules on the substrate at room temperature. Infrared spectroscopy alone would be insufficient for identifying individual molecules; therefore, ex-situ analysis of the residues using a high-resolution mass spectrometer is planned for future work, and will be discussed in a separate article.

\begin{figure}[!ht]
    \centering
    \resizebox{\hsize}{!}{\includegraphics{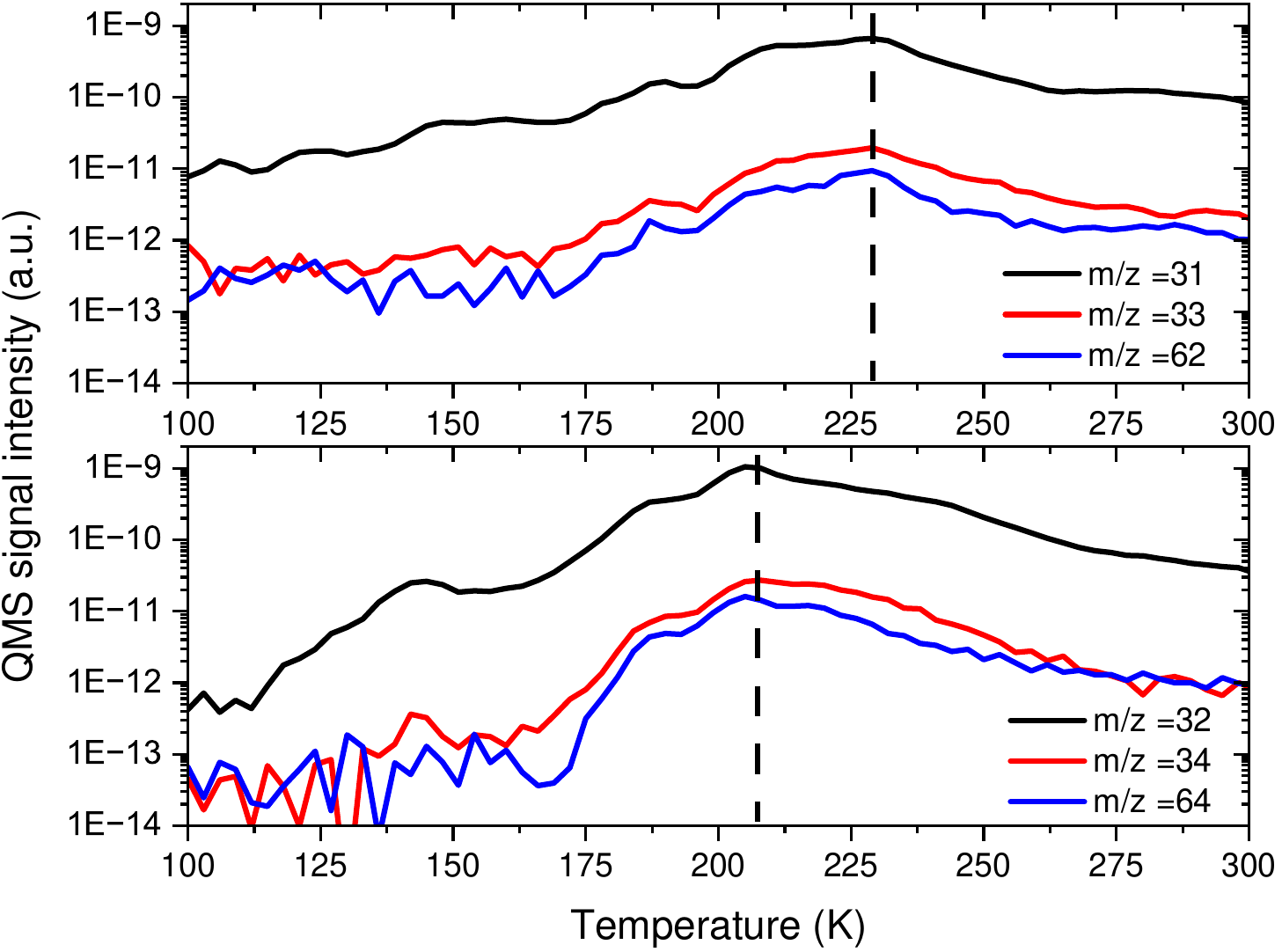}}
    \caption{Infrared spectra of the residues remaining after UV processing of about 60 ML and about 50 ML $^{12}$C EA on (a) KBr (b) Si, respectively, at room temperature. For comparison, the spectrum of about 60 ML $^{13}$C EA on (c) KBr is also presented. The spectra in (b) and (c) have been magnified 1.5 times for clarity.}
    \label{residue}
\end{figure}

The IR spectrum of the residue displays a broad band between 3000 and 3700 cm$^{-1}$, which can be attributed to N-H and O-H stretching bands. These broad bands are often associated with amines, alcohols, or carboxylic acids. The spectral region between 2550 and 3000 cm$^{-1}$ can be attributed to the aliphatic C-H bonds which are likely due to -CH$_3$ and -CH$_2$ containing species. No features are observed between 2300 and 2000 cm$^{-1}$, which are typically indicative of nitriles and isonitriles \citep{accolla2018}.

The region from 1800 to 950 cm$^{-1}$ is complex and challenging due to overlapping signals from various functional groups, including amides, carboxylic acids, ketones, and esters, as reported in the literature. Notably, after irradiation, we observe the infrared features of amide I, II, and III bands, which are also present in the residue. Recent studies confirm that these bands correspond to proteinogenic and non-proteinogenic amino acids, dipeptides, and even polypeptides \citep{kaiser2013}. We compared our residue analysis with selected studies available in the literature, though this comparison is not exhaustive. Detailed peak assignments and corresponding references are summarized in Table \ref{table3}. The amide peaks are also reminiscent of the peptide peaks described in a recent study by \citet{Krasnokutski2022, Krasnokutski2024}, where aminoketene was identified as a reaction intermediate, that could link EA to the formation of amino acids and peptides.

The infrared spectrum of the residue formed from UV irradiation of $^{13}$C-labeled EA on KBr is presented in Figure \ref{residue}. The key difference is observed in the Amide I and Amide II bands, where the $^{12}$C variant shows peaks at 1668 and 1579 cm$^{-1}$, which shift to 1618 and 1514 cm$^{-1}$, respectively. This shift clearly confirms that the amide groups originate from EA and are not due to any impurities. Additionally, there is a shift of approximately 15-20 cm$^{-1}$ in the bands at 1235 and 1053 cm$^{-1}$. This further underscores the accuracy of the band assignments, as they involve the stretching vibrations of C-N or C-O bonds. We would also like to point out an intriguing comparison with the spectrum in Fig. 4 of \citet{caro2003}, where residues formed by the simultaneous deposition and UV irradiation of formamide, followed by warming to room temperature, exhibit a remarkable similarity to the infrared spectrum of the residue at room temperature in the present study. Some similarities are expected as both compounds contain CHON elements and UV irradiation could lead to significant fragmentation of the parent molecules into smaller radicals. However, the two molecules have different functional groups to drive the chemistry. Further research is needed to explore these differences in more detail.

\startlongtable
\begin{deluxetable*}{cccc}
\tablecaption{Infrared absorption assignments of residue formed after irradiation and warming up of EA ice at room temperature.}
\label{table3}
\tablewidth{0pt}
\tablehead{
\colhead{Band position (cm$^{-1}$)} & \colhead{Assignment} & \colhead{Possible carrier species} & \colhead{Reference} 
}
\startdata
3226& $\nu(NH)$, $\nu(NH_2)$, $\nu(OH)$ & Amines, Alcohols, Carboxylic acids &  1, 2\\
2933 & $\nu_a(CH_2)$ & Aliphatic alcohols& 3\\
2856& $\nu_s(CH_3)$& Aliphatic alcohols & 3 \\
1668 & $\nu(C=O)$ &Amides, Amide I  & 4 \\
1579& $\nu_a(COO^-)$ &Carboxylic acids,  Amide II & 1, 4 \\
1465 & $\delta(NH)$, $\delta(CH_2)$, $\delta(CH_3)$ & Ammonium salts & 1\\
1366& $\delta(CH)$, $\nu_s(COO^-)$& carboxylic acids & 1, 4 \\
1235 & $\nu(CN)$, $\nu(O=C-O)$ & ketone &  1\\
1053 & $\nu(C-O)$ & Methyl ethers, Alcohols, polyoxymethylene like species  & 3, 5  \\
\hline 
\enddata
\tablerefs{\citet{caro2003}[1], \citet{accolla2018}[2], \citet{agarwal1985}[3], \citet{kaiser2013}[4]}, \citet{bernstein1995}[5]
\end{deluxetable*}

\section{Astrophysical implications}

It is important to note that under dense cloud conditions, water ice is the primary component of the icy mantles, and if EA is present, it will be diluted within the water ice. Therefore, it is essential to study EA mixed with water ice, especially in the context of energetic processing. The motivation for first studying pure EA ice is to better understand its role in prebiotic chemistry. The literature suggests that water ice can act as a shield to protect the other components from energetic processing and also potentially enhance the reaction network, making it more complex to interpret. Future studies will aim to further investigate this EA-water mixture.

The UV photon flux of 4.8$\times$10$^{3}$ cm$^{-2}$ s$^{-1}$, estimated for the attenuated field within a dense molecular cloud \citep{mennella2003}, and the UV photon flux of 8$\times$10$^{7}$ cm$^{-2}$ s$^{-1}$, calculated for the average interstellar radiation field in the diffuse medium \citep{mathis1983}, correspond to EA destruction rates of 2.3$\times$10$^{-14}$ s$^{-1}$ and 3.8$\times$10$^{-10}$ s$^{-1}$, respectively. Following \citep{mate2018}, we have also calculated the half-life time ($\mathrm{T_{1/2}}$) of EA using the following relations:

\begin{equation}
\label{B}
\mathrm{D_{1/2}} = \mathrm{F_{1/2}\times{Ef_E}}
,\end{equation}

\begin{equation}
\label{C}
\mathrm{T_{1/2}} = \mathrm{D_{1/2}/{r}}
\end{equation}

where, the quantity $\mathrm{F_{1/2}} = 2.34 \times 10^{17}$ photons cm$^{-2}$ represents the half-life fluence, which can be estimated from Equation \ref{A} by setting $N(t) = 0.5$ and using the derived values for $\sigma_d$ and $a$. $\mathrm{E}$ denotes the photon energy in units of eV molecule$^{-1}$ cm$^{2}$ and $\mathrm{f_E}$ is the fraction of energy absorbed (estimated above), and $\mathrm{r}$ is the UV dose rate in the space environment of interest. Based on this derivation, we estimate $\mathrm{D_{1/2}}$ = 32.2 eV molecule$^{-1}$ and $\mathrm{T_{1/2}}$ = $6.5\times10^{7}$ yr for EA ice in cold dense interstellar clouds. UV dose rates in these clouds are $4\times10^{-7}$ eV molecule$^{-1}$ yr$^{-1}$\citep{moore2001}. This half-life is comparable to the typical lifetime of a dense cloud (around 10$^7$ yr) and is of a similar order of magnitude as the estimated half-life of pure EA ice exposed to 1 keV electron irradiation (simulating the effects of cosmic rays) \citep{zhang2024}. The results of the half-life estimate are significant as they indicate that EA ice is capable of resisting UV destruction and remaining intact throughout the entire lifespan of the dense cloud, thereby allowing sufficient time for chemical reactions to produce prebiotic COMs.

\begin{figure*}
    \centering
    \resizebox{\hsize}{!}{\includegraphics{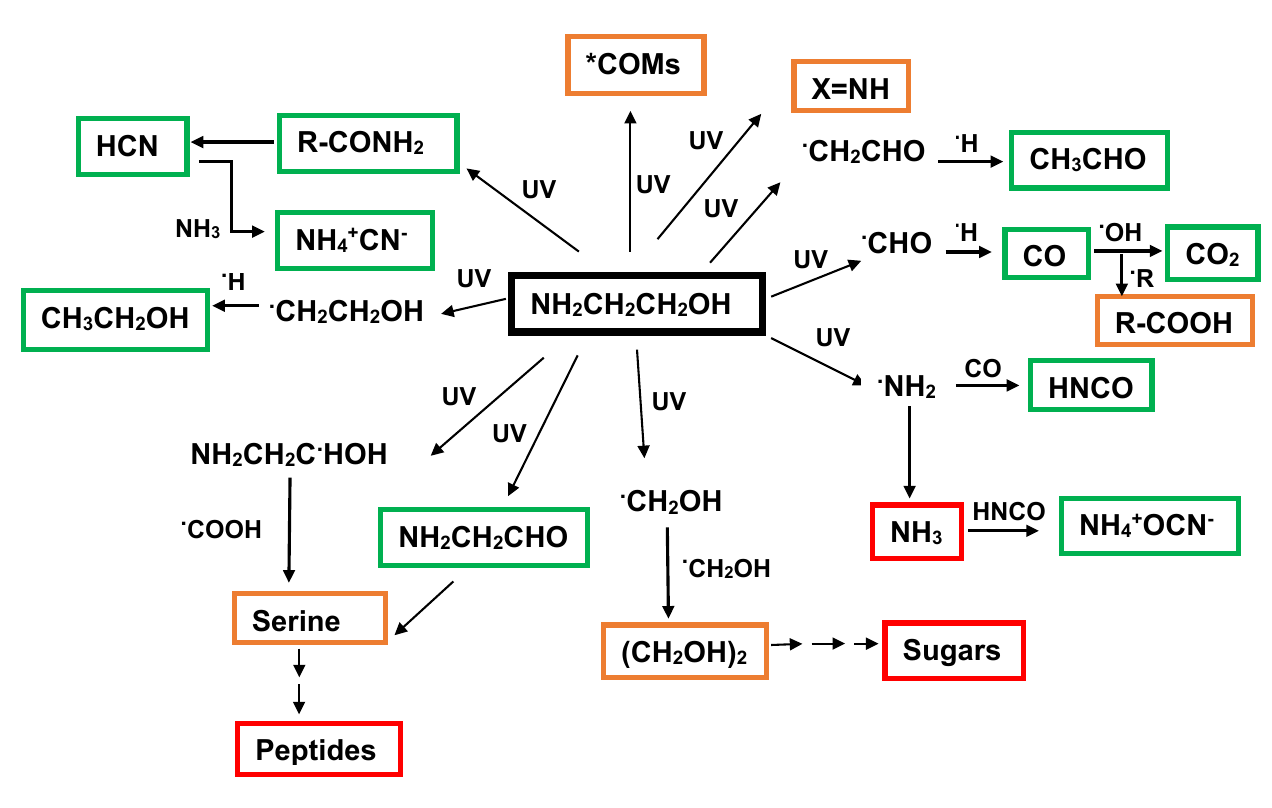}}
    \caption{Summary of the reaction scheme showing the formation of new products after UV irradiation of EA. The green, orange, and red boxes indicate detected, tentative, and undetected products. The successive arrows indicate there could be more than one step involved towards the final product.COMs with an asterix symbol indicates more complex ogranic molecules could be there but not yet confirmed.}
    \label{Scheme}
\end{figure*}
\clearpage

\section{Conclusions}

This study investigates the photostability of EA ice under Ly-$\alpha$ irradiation at 10 K in high vacuum conditions, mimicking the environment of dense molecular clouds. The study led to the estimation of the destruction cross-section and half-life of EA under these conditions. UV photolysis produced both simple and complex species, tentative identifications include ethylene glycol and serine, suggesting that EA could serve as a precursor to prebiotic molecules such as amino acids and sugar derivatives in the interstellar environment. Future research will focus on analysing the non-volatile residues left after irradiation, which may reveal additional complex organic compounds relevant to the origin of life.

\begin{acknowledgments}
The authors are grateful to G. Rouill{\'e} for his technical support and the evaluation of the Ly-$\alpha$ photon flux. This work has been supported by the European Research Council under the Horizon 2020 Framework Program via the ERC Advanced Grant Origins 83 24 28. We would also like to acknowledge funding through the NanoSpace COST action (CA21126 – European Cooperation in Science and Technology) and Vector Stiftung (Project ID P2023-0152). SAK is grateful to DFG (grant \textnumero \; KR 3995/4-2). 
\end{acknowledgments}

\bibliography{REF}{}
\bibliographystyle{aasjournal}

\end{document}